\documentstyle[12pt]{article}
\def\pzv#1#2{{\partial{#1}\over\partial{#2}}}
\begin{document}
\title{An asymmetric double-slit interferometer for small and large
quantum particles}
\author{Mirjana Bo\v zi\'c$^1$, Du\v san Arsenovi\'c$^1$ and Lep\v sa Vu\v skovi\'c$^2$\\
$^1$Institute of Physics, P.O.\ Box 57,\\
11001 Belgrade, Serbia and Montenegro\\
e-mail: bozic@phy.bg.ac.yu, arsenovic@phy.bg.ac.yu\\
$^2$Old Dominion University, Department of Physics,\\
4600 Elkhorn Avenue, Norfolk, VA 23529, USA\\
e-mail: vuskovic@physics.odu.edu}
\maketitle
\vfill
\newpage
\begin{abstract}
Quantum theory of interference phenomena does not take the diameter of 
the particle into account, since particles were much smaller than the width
of the slits in early observations. In recent experiments with large
molecules, the diameter of the particle has approached the width of the slits.
Therefore, analytical description of these cases should include a
finite particle size. The generic quantum interference setup is an
asymmetric double slit interferometer. We evaluate the wave
function of the particle transverse motion using two forms of the
solution of Schr\"odinger's equation in an asymmetric interferometer:
the Fresnel-Kirchhoff form and the form derived from the transverse
wave function in the momentum representation. The transverse momentum
distribution is independent of the distance from the slits, while the
space distribution strongly depends on this distance. Based on the
transverse momentum distribution we determined the space distribution
of particles behind the slits. We will present two cases: {\it a\/})
when the diameter of the particle may be neglected with respect to the width
of both slits, and {\it b\/}) when the diameter of the particle is larger than
the width of the smaller slit.
\newpage
\end{abstract}
\section{Introduction}
\null\hskip\parindent
Until recently various quantum interference experiments were conducted
with objects (photons, electrons, neutrons,...) of the size
much smaller than the characteristic dimensions of the
diffraction structure \cite{Martini}. The first single slit experiment with
Rydberg atoms, objects of non negligible size with respect 
to the width ot the slit was performed by Fabre et al. \cite{Fabre}. By 
measuring  transmission through  micrometer size slits, these authors
determined the size of Rydberg atoms. Later,  Hunter and Wadlinger 
\cite{Hunter}  proposed the single-slit diffraction experiment in order to 
measure the diameter of the photon. 

In order to investigate the reasons of unobservability of quantum
effects in the classical world Arndt et al. \cite{Arndt}, Nairz et al.
\cite{Nairz}, Brezger et al. \cite{Brezger} performed quantum
interference experiments with objects of large mass and diameter, including
macromolecules. The experiments raise various questions 
about theoretical concepts. The following three are obvious: 

1) Are there new effects applying to particles bigger than their de
Broglie wavelength?

2) Does internal structure have influence on interference?

Arndt et.\ al.\ emphasized the task \cite{Arndt}: ``Here we report the
observation of de Broglie wave interference of ${\rm C}_{60}$ molecules
by  diffraction at a material absorption grating. This molecule is the
most massive and complex object in which wave behavior has been
observed. Of particular interest is the fact that ${\rm C}_{60}$ is
almost a classical body, because of its many excited internal degrees
of freedom...''.

3) What happened with an ensemble of incoming particles if slits are
smaller than the diameter of the particles?

In the experiment of Arndt et.\ al.\ the de Broglie wavelength of the
interfering fullerenes is already smaller than their diameter by a
factor of almost $400$ and authors pointed out that ``it would be
certainly interesting to investigate the interference of objects the
size of which is equal or even bigger than the diffracting structure''
\cite{Arndt}.

These experiments could shed more light on a long standing dilema
whether each quanton consists of a particle and accompanied wave (as
two different compatible entities) \cite{deBroglie}, or quantons
sometimes behave like a wave and sometimes behave like a particle
(obeying principle of complementarity) \cite{Bohr}? The following
citations illustrate the present situation.

``We performed an experiment which was proposed by Ghose, Home and
Agarwal showing both classical wave-like and particle-like behaviors of
single photon states of light in a single experiment, in conformity
with quantum optics.'' \cite{Mizobuchi}. ``Here we give a detailed
justification of our claim that this experimental results contradict
the tenet of mutual exclusiveness of classical wave and particle
pictures assumed in Bohr's complementarity principle.'' \cite{Ghose}.
``Simultaneous observations of wave and particle behavior is
prohibited'' \cite{Buks}.

``Although interference patterns were once thought
of as evidence for wave motion, when looked at in detail it can be seen
that the electron arrive in individual lumps. ... We must therefore
conclude that electrons show wave-like interference in their arrival
pattern despite the fact that they arrive in lumps just like bullets''.
\cite{Hey}

``It is frequently said or implied that the wave-particle duality of
matter embodies the notion that a particle -- the electron, for example
-- propagates like a wave, but registers at a detector like a particle.
Here one must again exercise care in expression so that what is already
intrinsically difficult to understand is not made more so by semantic
confusion. The manifestations of wave-like behavior are statistical in
nature and {\em always} emerge from the collective outcome of many
electron events... That electrons behave singly as particles and
collectively as waves is indeed mysterious, ... '' \cite{Silverman}

``Each atom is therefore at the same time a particle and a wave, the wave
allowing one to get the probability to observe the particle at a given
place.'' \cite{Cohen-Tannoudji} 

``...Ever since then the two sides of the same quantum object appeared
together: on the one hand the non-local wave nature needed to describe
the unperturbed propagation and on the other hand the local aspect of
the object when it is registered by the detector" \cite{Arndt}

In this paper we present the theoretical study of the dependence of the
quantum interference pattern on the diameter of the particle, assuming 
that the characteristic sizes of the diffraction structure 
are of the order of the diameter of the particle. We argue that an
asymmetric double-slit interferometer (an interferometer whose slits
have different widths $\delta_1$ and $\delta_2$) is the generic case
for this study.

{\raggedright
\section{The particle wave function behind an asymmetric grating}
}
\null\hskip\parindent 
We shall now determine, the wave function of a quanton which travels
with velocity, $\vec v=v \vec i$ through the region I, towards the slits and is
then sent through the slits to the region II (Fig.~1). Results in this
section are valid for arbitrary slits. This wave
function is a stationary solution of the time-dependent two dimensional
Schr\"odinger equation
\begin{equation}
-{\hbar^2\over2m}\left(\pzv{^2}{x^2}+\pzv{^2}{y^2}\right)\Psi(x,y,t)=i\hbar\pzv{}t\Psi(x,y,t).\label{timeschr}
\end{equation}
The solution of (\ref{timeschr}) has the form
\begin{equation}
\Psi(x,y,t)=e^{-i\omega t}\varphi(x,y),\label{two}
\end{equation}
where $\hbar\omega=mv^2/2$ and $p=mv=\hbar k$. Space dependent
function $\varphi(x,y)$ satisfies the Helmholtz equation
\begin{equation}
-{\hbar^2\over2m}\left(\pzv{^2}{x^2}+\pzv{^2}{y^2}\right)\varphi(x,y)=\hbar\omega\varphi(x,y).\label{statschr}
\end{equation}
The solution of this equation in the region I is a spherical wave
\begin{equation}
\varphi(P^\prime)=\varphi(x^\prime,y^\prime)=A{e^{ikr^\prime}\over r^\prime},
\end{equation}
where $A$ is a constant and $r^\prime$ is a distance (Fig.~1) from the source
($P_0$) to the point $P^\prime=(x^\prime,y^\prime)$ in the region I.
The distance $a$ of the double-slit screen from the source $P_0$ being
very large compared to the width of the slits, this spherical wave at
the slit points $(x^\prime=x^{\prime\prime},y^\prime=0)$ may be approximated by the
plane wave. In the region II the equation (\ref{statschr}) is as simple
as before but initial condition makes the solution more difficult.

Solution known as Fresnel-Kirchhoff diffraction formula \cite{BornWolf} reads:
\begin{equation}
\varphi(x,y)=-{iA\over2\lambda}{e^{ika}\over a}\int_{\cal
A}dx^{\prime\prime}{e^{iks}\over s}[1+\cos\chi],\label{fksol}
\end{equation}
where $s=\sqrt{y^2+(x^{\prime\prime}-x)^2}$, $\cos\chi=y/s$,
$\lambda=2\pi/k$.  The region $\cal A$ is the union of all intervals
along the $x$-axis where slits are open. From now on $x^{\prime\prime}$
represents a variable of integration along the line of the slits.

Far enough from the slits wave function resembles the Fourier transform
of the wave field accross the aperture. This can be verified from the
Fresnel-Kirchhoff solution. In the far region it follows:
\begin{equation}
\varphi(x,y)\approx-{iA\over\lambda}{e^{ika}\over a}{e^{iky}\over
y}\int_{\cal A}dx^{\prime\prime}\,\varphi(x^{\prime\prime},0)\,e^{-ikxx^{\prime\prime}/y}.
\end{equation}
Wave function is now separable into two functions, one depending on $y$ and
the other depending on $K_x\equiv kx/y$ \cite{Hecht}:
\begin{eqnarray}
&\displaystyle\varphi(x,y)=D(y){\cal F}(K_x)&\nonumber\\
&\displaystyle D(y)=-\sqrt{2\pi}{iA\over\lambda}{e^{ika}\over a}{e^{iky}\over
y}&\nonumber\\
&\displaystyle{\cal F}(K_x)={1\over\sqrt{2\pi}}\int_{\cal A}dx^{\prime\prime}\,\varphi(x^{\prime\prime},0)\,e^{-iK_xx^{\prime\prime}}.&\label{faraway}
\end{eqnarray}

The solution of equation (\ref{statschr}) in region II can be
written in another
form \cite{Naturforsch,Arsenovic,Vuskovic}. This form is more convenient for our
analysis than the form (\ref{fksol}).
With approximation valid for small diffraction angles $\chi$ we have:
\begin{equation}
\varphi(x,y)=e^{iky}{1\over\sqrt{2\pi}}\int_{-\infty}^{+\infty}dk_x\,c(k_x)\,e^{ik_xx}e^{-i{k_x^2\over2k}y}\equiv
e^{iky}\phi(x,y).\label{planewave}
\end{equation}
where $c(k_x)$ is the Fourier transform of the function $\varphi(x,y)$
on the aperture $\varphi(x,y=0)$:
\begin{equation}
c(k_x)={1\over\sqrt{2\pi}}\int_{-\infty}^{+\infty}dx^{\prime\prime}\,\varphi(x^{\prime\prime},0)\,e^{-ik_xx^{\prime\prime}}.\label{cdef}
\end{equation}
Inserting (\ref{cdef}) into (\ref{planewave}), after
integration over $k_x$ one finds
\begin{equation}
\varphi(x,y)=e^{-i{\pi\over4}}e^{iky}\sqrt{k\over2\pi}{1\over\sqrt
y}\int_{\cal
A}\varphi(x^{\prime\prime},0)e^{i{k(x-x^{\prime\prime})^2\over2y}}dx^{\prime\prime}.\label{plwa}
\end{equation}
This function is normalized
$\int_{-\infty}^{+\infty}\vert\varphi(x^{\prime\prime},y)\vert^2\,dx^{\prime\prime}=1$, provided
$\int_{\cal A}\vert\varphi(x^{\prime\prime},0)\vert^2\allowbreak dx^{\prime\prime}=1$.
The form (\ref{plwa}) clearly expresses wave
function's dependence on the boundary condition and it is appropriate for
numerical computation.

For large values of $y$ the function $\varphi(x,y)$ in (\ref{plwa}) is
approximated by
\begin{equation}
\varphi(x,y)=\sqrt{k\over2\pi y}e^{-i{\pi\over4}}e^{iky}e^{ikx^2/2y}\int_{\cal A}\varphi(x^{\prime\prime},0)e^{-ikxx^{\prime\prime}/y}dx^{\prime\prime}.\label{plwaa}
\end{equation}
Taking Eq.~(\ref{cdef}) into account, Eq.~(\ref{plwaa}) takes the form
\begin{equation}
\varphi(x,y)=e^{iky}\sqrt{k\over y}e^{-i{\pi\over4}}e^{ikx^2/2y}c(kx/y).\label{farc}
\end{equation}
We see that the variable $K_x={kx\over y}={mx\over\hbar t}$ plays the role of
$k_x$.

Since $K_x$ is proportional to $x/y$ functions $\vert\varphi(x,y)=const\vert$ 
are family of functions of $x$ spreading along the  $x-$ axis as $y$ increases. 
In fact, for each value of $\vert\varphi\vert$, in the far field there
exists the straight line with origin at the center of the grating along
which this particular value of $\vert\varphi\vert$ propagates.

{\raggedright
\section{The understanding of the space distribution using transverse
momentum distribution}
}
\null\hskip\parindent
By assuming that the motion of an atom along the $y$-axis may be
treated classically and that the transverse motion is quantum, one is
tempted to use the relation $y=vt$ and to determine the time dependent
function of the transverse motion $\psi(x,t)$ from the function
$\phi(x,y)$, by the following definition:
\begin{equation}
\psi(x,t)\equiv\phi(x,vt)={1\over\sqrt{2\pi}}\int dk_x\,c(k_x)e^{ik_xx}e^{-i\omega_xt}\label{ft}
\end{equation}
where $\omega_x=\hbar k_x^2/2m$. We see that the function $\psi(x,t)$
has the form of a general solution of the one-dimensional Schr\"odinger
equation. The wave function $c(k_x)$ is then seen as a wave function of
this one-dimensional (transverse) motion in the momentum
representation. It's modulus square, $\vert c(k_x)\vert^2$, determines
the distribution of transverse momenta. The wave function $\Psi(x,y,t)$
from Eq.~(\ref{two}) is expressed through $\psi(x,t)$ as
\begin{equation}
\Psi(x,y,t)=e^{iky}e^{-i\omega t}\psi(x,t).
\end{equation}

By taking Eq.~(\ref{farc}) into account one concludes that in the far field
the relation (\ref{ft}) between the wave functions $\psi(x,t)$
and $c(k_x)$ reduces to the simplier form:
\begin{equation}
\psi\left(x,t={ym\over\hbar k}\right)=\sqrt{k\over y}e^{-i{\pi\over4}}e^{ikx^2/2y}c(kx/y).
\end{equation}
Based on the above factorization of the wave function $\Psi(x,y,t)$ and
the properties of its factors summarized above, we proposed
\cite{Arsenovic} the new expression for the probability density $\tilde
P(x,t)$ fo the particle's arrival to a certain point $(x,y=vt)$ at time
$t$:
\begin{eqnarray}
&\displaystyle\tilde P\left(x,{y\over v}\right)=\tilde
P(x,t)\equiv&\nonumber\\
&\displaystyle\equiv\int_{-\infty}^{+\infty}dk_x\int_{-\infty}^{+\infty}dx^{\prime\prime}\,\vert
c_n(k_x)\vert^2\vert\phi(x^{\prime\prime},0)\vert^2\delta\left(x-x^{\prime\prime}-{\hbar
k_xt\over m}\right).&\label{trajec}
\end{eqnarray}
Particles emerge from different points $(x^{\prime\prime},0)$ at the aperture. 
That is the reason for integration over $x^{\prime\prime}$. The
contribution of each point at the aperture  is proportional to
$\vert\phi(x^{\prime\prime},0)\vert^2$. The integration over
$dk_x$ and the function $\vert c_n(k_x)\vert^2$ reflect the contribution
of various angles/momenta in diffraction. Finally, $\delta$-function assumes
straight trajectory from a point $(x^{\prime\prime},0)$ at the slits to
the point $(x,y)$ and leads to the simplified form 
\begin{equation}
\tilde P(x,t)=\int_{-\infty}^{+\infty}dk_x\,\vert
c_n(k_x)\vert^2\left\vert\phi\left(x-{\hbar k_xt\over m},0\right)\right\vert^2.\label{sumtraj}
\end{equation}

By assuming that the function $\phi(x^{\prime\prime},0)=0$ for
$x^{\prime\prime}\not\in{\cal A}$ and
$\phi(x^{\prime\prime},0)=const$ such that $\int_{\cal
A}\vert\phi(x^{\prime\prime},0)\vert^2=1$, for
$x^{\prime\prime}\in{\cal A}$, the Eq.~(\ref{sumtraj}) is transformed
to the following usefull form
\begin{equation}
\tilde
P(x,t)={1\over\sqrt{\sum_{i=1}^n\delta_i}}\sum_{i=1}^n\int_{{m\over\hbar}(x-x_r^i)}^{{m\over\hbar}(x-x_l^i)}dk_x\vert
c(k_x)\vert^2\equiv\sum_{i=1}^n\tilde P_i(x,t).
\label{sumck}
\end{equation}
Here $x_l^i$ and $x_r^i$ are the coordinates of the left and right edge
of the $i$-th slit.

The total probability density $\tilde P(x,t)$ is a sum of $n$ terms,
$\tilde P_i(x,t)$. $\tilde P_i(x,t)$ is interpreted to be the
probability that a quanton reaches $(x,y=vt)$ at time $t$ after passing
through the $i$-th slit of the $n$-slits grating.

Numerical calculation shows that far from the slits the function $\tilde
P(x,t)$ (Fig.~4) is very nearly equal to the exact probability density
$\vert\Psi(x,y,t)\vert^2=\vert\psi(x,t)\vert^2$ (Fig.~2). Near the
slits $\tilde P(x,t)$ and $\vert\psi(x,t)\vert^2$ qualitatively look
similarly but they differ numerically.

{\raggedright
\section{On the possible influence of particle's diameter on the
interference pattern}
}
\null\hskip\parindent
We outline an approach to investigate how the widths of the slits influence the
interference pattern in the double-slit experiment with quantons -
photons, electrons, neutrons, atoms, molecules.

Interference effects are visible when the wavelength of quantons is of
the order of the distance between the slits. In practice, this distance
is $d=(2-50)\,\lambda$. The slit width is often equal or up to ten
times smaller than the distance between the slits. In quantum
interference experiments with electrons and neutrons the diameter of
the particle is smaller than the wavelength. Consequently, in classical
experiments the width $\delta$ of the slits is much greater than
$D$. But, depending on the velocity, atoms may have de Broglie
wavelength which is smaller than the diameter of the atom. With macromolecules
such a situation encounters more often, as shown in the experiment of
Arndt et.\ al.\ and discussed by Arndt et al.\ \cite{Arndt} and Nairz
et al.\ \cite{Nairz}. So,
interference experiments with such quantum particles could have the
slit widths smaller than the particle diameter.

This requires a theoretical approach to quantum interference which
takes the diameter of the particle into account \cite{Bozic}. A study of a quantum particle in
an asymmetric double-slit interferometer ($\delta_1>\delta_2$) seems to
be useful for this purpose because we identify two characteristic cases
for the ratio of slit widths $\delta_1$ and $\delta_2$ and the diameter
of the particle $D$:

{\it a\/}) The diameter $D$ is negligeable with respect to the
widths $\delta_1$ and $\delta_2$.

{\it b\/}) The diameter $D$ is greater than the width $\delta_2$,
where $\delta_2<\delta_1$.

In the case {\it a\/}), which was until recently the only case of 
physical interest, there is no need to consider or  take into 
account the diameter of the particle. The particle momentum $\vert
c(k_x)\vert^2$ and space distribution $\vert\psi(x,t)\vert^2$ behind
the grating are determined by the wave function
\begin{equation}
\psi(x,t)=\phi(x,vt)={1\over\sqrt{2\pi}}\int_{-\infty}^{+\infty}c(k_x)e^{i(k_xx-\omega_x
t)}dk_x
\label{ift}
\end{equation}
where
\begin{equation}
\phi(x,0)=\psi(x,0)=\cases{{1\over\sqrt{\delta_1+\delta_2}}&$x\in{\cal
A},\,\,\,{\cal A}=\left({-d-\delta_1\over2},{-d+\delta_1\over2}\right)\cup\left({d-\delta_2\over2},{d+\delta_2\over2}\right)$\cr
\strut0&$x\not\in{\cal A}$\cr}
\label{cases}
\end{equation}
and
\begin{equation}
c(k_x)={1\over\sqrt{2\pi(\delta_1+\delta_2)}}{2\over k_x}\left[e^{ik_xd/2}\sin{k_x\delta_1\over2}+e^{-ik_xd/2}\sin{k_x\delta_2\over2}\right].\label{cktwo}
\label{cdefd}
\end{equation}
The functions $\vert\psi(x,t)\vert^2$ and $\vert c(k_x)\vert^2$ are
graphically represented at Fig.~2 and Fig.~3, for the chosen set of
parameters.

In the case {\it b\/}), we are faced with the question how and where to
take the diameter of the particle into account. We know that the diameter
of the particle is not incorporated anywhere in the Schr\"odinger equation. 
But, we expect that a particle with diameter $D$, such that
$\delta_1>D>\delta_2$ could not pass through the second slit.

So, it seems to us that we are forced to assume that wave functions in
the coordinate and momentum representation in the case {\it b\/}) is
also given by expressions (\ref{ift})-(\ref{cktwo}).

The momentum distribution $\vert c(k_x)\vert^2$ of particles is given also
by (\ref{cdefd}), because it is determined by the values of the wave
function at the boundary.

But the space distribution of particles in case {\it b\/}) is different
from the space distribution in case {\it a\/}), because the particles
arriving to the smaller slit can not go through. We conclude that
particle distribution in case {\it b\/}) is given by $\tilde P_1(x,t)$
from the expression (\ref{sumck}) of $\tilde P(x,t)$.
\begin{equation}
\tilde P(x,t)\approx\tilde
P_1(x,t)={1\over\sqrt{\sum_{i=1}^n\delta_i}}\int_{{m\over\hbar}(x-x_r^i)}^{{m\over\hbar}(x-x_l^i)}\vert
c(k_x)\vert^2dk_x.
\end{equation}
The probability $\tilde P_1(x,t)$ is graphically represented in Fig.\
5.
\section{Conclusion}
Inspired by current efforts to perform diffraction and interference
experiments with objects of size that is equal or even larger than the
diffraction structure, we outline an approach to investigate how the
particle diameter influences the interference pattern in an asymmetric
double slit interferometer.

We identify two characteristic cases for the ratio of slit widths
$\delta_1$ and $\delta_2$ and the diameter $D$ of the particle: {\it
a\/}) $D\ll\delta_1$ and $D\ll\delta_2$, {\it b\/})
$\delta_1>D>\delta_2$. The wave function behind the grating has the
same form in both cases because it is the solution of the Schr\"odinger
equation which is not sensitive to the diameter of the particle.

The space distribution of particles in case {\it a\/}) is given as
usual by the modulus square of this function. Using the same wave
function and assuming that a particle with diameter $D$, such that
$\delta_1>D>\delta_2$ could not pass through the second slit, we
determine the space distribution in case {\it b\/}). We conclude that
the momentum distribution of particles behind the grating is the same
in cases {\it a\/}) and {\it b\/}).
\null\hskip\parindent

\newpage
\centerline{\bf Figure captions}
\vskip\baselineskip
Fig.\ 1. Illustration of a grating with $n$ slits of various widths.

Fig.\ 2. The particle distribution function $\vert\psi(x,t)\vert^2$
behind the asymmetric double slit grating ($\delta_1=1\,\mu{\rm m}$,
$\delta_2=0.25\,\mu{\rm m}$, $d=8\,\mu{\rm m}$) close to the slits
$(a,b)$ and far from the slits $(c,d)$. It is evaluated from the form
$(19)$ of the wave function. The initial longitudinal wave vector is
$k=4\pi\cdot10^{10}\,{\rm m}^{-1}$, the particle mass is
$m=3.8189\cdot10^{-26}\,{\rm kg}$.

Fig.\ 3. The particle transverse momentum distribution $\vert
c(k_x)\vert^2$ behind the asymmetric double-slit grating
($\delta_1=1\,\mu{\rm m}$, $\delta_2=0.25\,\mu{\rm m}$, $d=8\,\mu{\rm m}$).

Fig.\ 4. The probability density $\tilde P(x,t)$ of particles arrival
to the point $x$ at time $t$ ($y=vt$) behind the asymmetric double slit
grating ($\delta_1=1\,\mu{\rm m}$, $\delta_2=0.25\,\mu{\rm m}$,
$d=8\,\mu{\rm m}$) close to the slits $(a,b)$ and far from the slits
$(c,d)$. It is evaluated from Eq.~(18). Particles' diameter $D$ is
negligible with respect to the widths of the slits. The initial
longitudinal wave vector is $k=4\pi\cdot10^{10}\,{\rm m}^{-1}$, the
particle mass is $m=3.8189\cdot10^{-26}\,{\rm kg}$.

Fig.\ 5. The probability density $\tilde P_1(x,t)$ of particles
reaching $(x,y)$ at time $t$ after passing through the larger slit,
near the slits $(a,b)$ and far from the slits $(c,d)$. It is evaluated
from Eq.~(22). $D$ is assumed to be larger than $\delta_2$ and smaller
than $\delta_1$. The values of parameters are the same as in captions
of Figs.~2,3,4.
\end{document}